\documentclass{article}
\usepackage{spconf,amsmath,graphicx}
\usepackage{hyperref}
\usepackage{blindtext}
 
\title{An End-to-End Neural Network for Image-to-Audio Transformation}
\name{Liu Chen$^{1,2}$, Michael Deisher$^2$, Munir Georges$^{3,4}$}

\address{
  $^1$Oregon Health and Science University, Oregon USA\\
  $^2$Intel Corporation, Hillsboro, Oregon USA\\
  $^3$Intel Labs, Munich, Germany\\
  $^4$Technische Hochschule Ingolstadt\\
  \href{mailto:chliu@ohsu.edu,michael.deisher@intel.com,munir.georges@intel.com}{chliu@ohsu.edu,\{michael.deisher,munir.georges\}@intel.com}}

\begin{document}
%\ninept
%
\maketitle
\begin{abstract}
This paper describes an end-to-end (E2E) neural architecture for the audio rendering of small portions of display content on low resource personal computing devices.  It is intended to address the problem of accessibility for vision-impaired or vision-distracted users at the hardware level.  Neural image-to-text (ITT) and text-to-speech (TTS) approaches are reviewed and a new technique is introduced to efficiently integrate them in a way that is both efficient and back-propagate-able, leading to a non-autoregressive E2E image-to-speech (ITS) neural network that is efficient and trainable.  Experimental results are presented showing that, compared with the non-E2E approach, the proposed E2E system is 29\% faster and uses 19\% fewer parameters with a 2\% reduction in phone accuracy.  A future direction to address accuracy is presented. 
\end{abstract}
\begin{keywords}
OCR, TTS, image-to-speech
\end{keywords}
\section{Introduction}
\label{sec:intro} 
Users of touchscreen-enabled personal computing devices such as cell phones, tablets, and laptops may encounter situations where safety considerations or visual impairment make it difficult to take in display content.  Operating system (OS) accessibility features that read aloud the text on the display can be used to mitigate this problem.  However, screen readers are typically subordinate to the OS and may not render text content within images.  The use of a dedicated neural network co-processor to implement such a capability has the advantages of low cost per watt, low power consumption, robustness to operating system failure, and application independence.  Although it is possible to generate audio from a region of the display using separate image recognition and audio production neural networks, a non-autoregressive E2E neural network architecture is more suitable for this application since it simplifies the hardware design and the inference procedure.  To keep power consumption and cost low it is also necessary to minimize the required memory footprint.  In this paper we introduce a non-autoregressive E2E neural network suitable for embedded hardware implementation of an image-to-speech subsystem in personal computing devices.  We build upon previous research in the areas of image-to-text (ITT) and text-to-speech (TTS) and introduce a novel method to bridge these into a single trainable ITS neural network.  As far as we are aware, this is the first time this has been done.

\begin{figure}[t!]
    \centering
    \includegraphics[width=\linewidth]{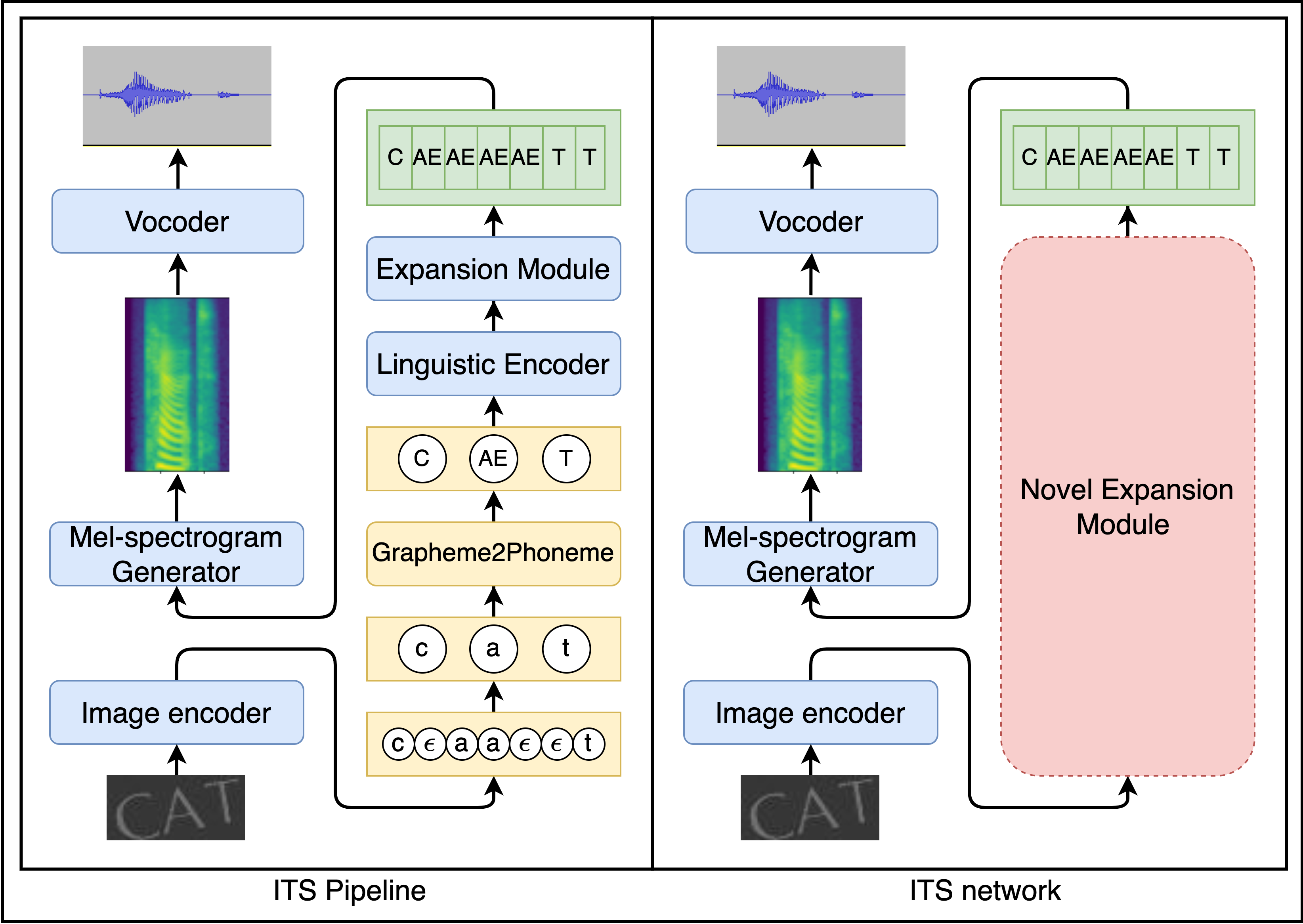}
    \caption{Image-to-Audio: non-back-propagateable post-/pre-processes steps in yellow are replaced by our novel expansion module shown in red. This enables end-to-end training.}
    \label{fig:non-e2e_vs_e2e}
\end{figure}

The network structures shown in Figure~\ref{fig:non-e2e_vs_e2e} synthesize speech given the fixed size image of a word. In the left half of the figure, we show the pipeline (detailed in Section \ref{sec:BACKGROUND}) using separate non-autoregressive ITT and TTS models. We desire a fully backpropagatable non-autoregressive E2E network.  However, the post-/pre-processing steps shown in yellow are non-back-propagatable. Therefore, we devote our efforts to the pink box in the right half of Figure~\ref{fig:non-e2e_vs_e2e} which addresses alignment of the speech with the encoded image. Our contribution is an intuitive sequence expansion module for the E2E ITS system that is flexible enough to accommodate a wide variety of image encoders and Mel-spectrogram generators.

The rest of the paper is organized as follows: Section~\ref{sec:BACKGROUND} describes the ITS problem in light of previous work.  Section~\ref{sec:ARCHITECTURE} presents the E2E DNN architecture in detail. Section~\ref{sec:experiments} introduces the training process and evaluation metrics. Finally, Section~\ref{sec:results} presents results followed by conclusions.

\section{Background}
\label{sec:BACKGROUND}
\textbf{Image-to-text (ITT)} aims to recognize text in input images. Generally, ITT contains three modules: an optional rectifier, an image encoder, and a sequential decoder. The rectifier segments and normalizes images through transforming various types of text irregularities in a unified way~\cite{shi2016robust}. The image encoder extracts hidden representations from the normalized image~\cite{shi2016robust}. And the decoder generates a sequence of characters based on the hidden representations. Non-autoregressive ITT predicts sequences of arbitrary length. In the training process, groundtruth text is expanded with additional placeholders, i.e. $\epsilon$ or repeated characters, making the expanded groundtruth have the same length as the model's hidden representation sequence before feeding it to the chosen loss function~\cite{graves2006connectionist,Jaderberg14c,Jaderberg16}. During inference, characters are obtained after removing the placeholders.  The expansion algorithms vary among loss functions.  Connectionist temporal classification (CTC)~\cite{graves2006connectionist} utilizes both repeated characters and $\epsilon$.  Aggregation cross-entropy~\cite{xie2019aggregation} only utilizes $\epsilon$ as the placeholder. Both loss functions minimize the distance between the target text and all possible expanded texts, thus, a placeholder can occur at any position in the raw output.  Cai et al.~\cite{cai2021revisiting} revises a traditional method: expanding groundtruth by inserting  $\epsilon$ at the tail and minimizing this expansion through cross-entropy (CE) loss. 
Cai et al.~\cite{cai2021revisiting} indicates that, with proper configuration, this method can achieve comparable performance to CTC~\cite{graves2006connectionist}. Therefore, we adopt this approach here.

\textbf{Text-to-speech (TTS)}, which aims to synthesize natural and intelligible speech given text, first converts the text (i.e., sequence of graphemes or phonemes) to acoustic features (e.g., sequence of Mel-spectra) and then transforms the acoustic features into audio samples through a vocoder. Since a phoneme sequence is much shorter than its Mel-spectra sequence, aligning phonemes with Mel frames is essential. Recent approaches to solving this problem include the use of autoregressive DNNs as well as the use of non-autoregressive DNNs which rely on the monotonicity of phoneme-to-Mel alignment and predict duration explicitly to bridge the length mismatch~\cite{ren2019fastspeech}.  These sequence expansion modules contain expansion length regulators as well as phoneme duration predictors for the hard alignment between a phoneme and its Mel frames. The ground-truth alignments are obtained from an external aligner. FastSpeech~\cite{ren2019fastspeech} and ParaNet~\cite{peng2020non}, which were the first proposed non-autoregressive models, utilize a pretrained autoregressive model to obtain the phoneme-level ground-truth alignments and FastSpeech2~\cite{ren2020fastspeech} utilizes a forced aligner for the same purpose. PortaSpeech~\cite{ren2021portaspeech} eases the negative effect of imprecise phoneme alignment by learning word-level hard alignment from external aligner and phoneme-level soft alignment through an attention mechanism. Parallel Tacotron2~\cite{elias2021parallel2} and ESA~\cite{donahue2020end} utilize a combination of differentiable duration predictor and attention-based length regulator to model the utterance duration and learn the phoneme alignment without an external aligner. Glow-TTS~\cite{kim2020glow} utilizes the invertible transformation property of flow-based models and searches the most probable monotonic alignments through dynamic programming. In this work, we adopt non-autoregressive TTS with phoneme sequence input.  We avoid grapheme sequence input due to problems reported in ~\cite{donahue2020end}.

\section{E2E Image-to-Speech Architecture}
\label{sec:ARCHITECTURE}

\begin{figure}
    \centering
    \includegraphics[width=\linewidth]{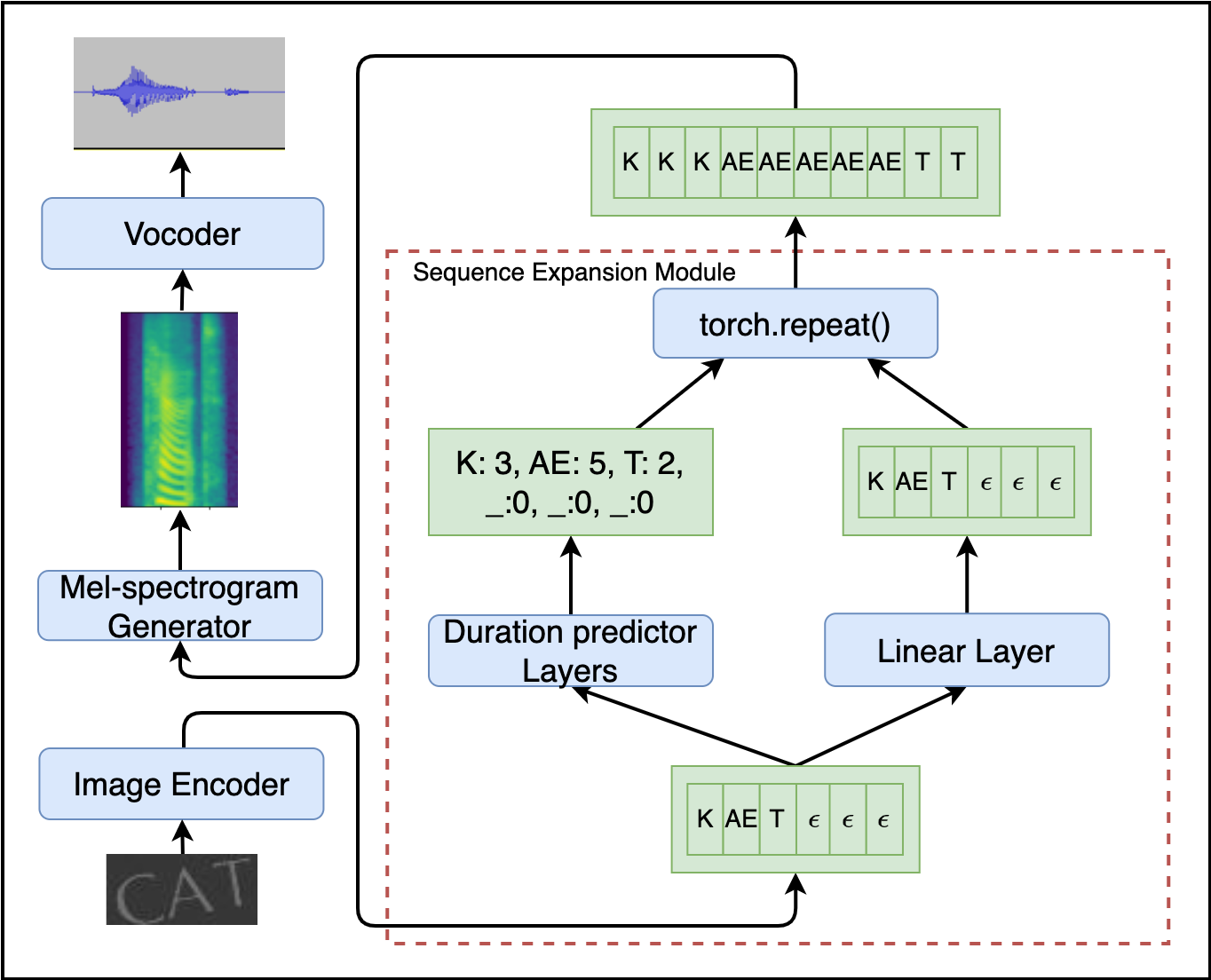}
    \caption{Processing an image containing the word "CAT". The symbol, $\epsilon$, represents a placeholder. The green rectangular boxes stand for hidden representations of symbols.}
    \label{fig:img2speech_general_pipeline}
\end{figure}

The core challenge of the E2E system is aligning Mel-spectrograms with the encoded image sequences with arbitrary length. 
In TTS, the length of each Mel-spectra sequence is derived directly from the encoded phoneme sequence.  However, in ITS, the lengths of encoded image sequences and the lengths of Mel-spectra sequences are not related. 
To bridge this gap, we introduce a non-phoneme symbol, $\epsilon$, as the only placeholder and expand phoneme sequences to an arbitrary fixed length by inserting $\epsilon$'s at the tail. 
Moreover, we define that only phonemes can be aligned to Mel-spectrograms and the duration of any $\epsilon$ must be zero. 
In summary, we assign two tasks to the duration predictor: 1) predicting phoneme durations (positive values) and 2) recognizing $\epsilon$'s (zeros). 
This eliminates the need for additional layers (e.g., used in~\cite{elias2021parallel}) for the second task.

We present our architecture with an example image in Figure~\ref{fig:img2speech_general_pipeline}.  First, the image encoder encodes ``cat'' into the hidden representation of the expanded phoneme sequence. Then the duration predictor predicts the duration of every vector in the sequence. 
The predictor generates zero if the hidden vector represents an $\epsilon$. 
Moreover, the Linear Layer on the right side transforms the hidden representation to the required dimension of Mel-spectra generator. 
Third, we expand the transformed representations through repeating each vector $d$ times where $d$ is its respective duration.
Fourth, the Mel-spectrogram generator takes the expanded representation as input and synthesizes the Mel-spectrogram.
Lastly, a vocoder synthesizes the raw waveform based on the synthesized Mel-spectrogram.

\smallskip\noindent\textbf{Image Encoder}. We follow the ITT architecture introduced in Section~\ref{sec:BACKGROUND}. We utilize the same rectifier as Shi et al~\cite{shi2016robust} and adopt HBONet~\cite{li2019hbonet} as the backbone network to extract hidden features from the rectified images. A pooling layer extracts global semantic information from hidden features and feeds it to 26 linear layers. The $i^{th}$ linear layer predicts the $i^{th}$ output. If a word only has $N$ phonemes, then the last {26-N} layers should predict $\epsilon$. Our configuration of HBONet is similar to~\cite{li2019hbonet} except we change the values of $n$ and $s$ of the first Inverted Residual block to be 1 (see Table 1 in ~\cite{li2019hbonet}). Empirically, we found that this modification  impacts the model accuracy very slightly while saving 0.5M parameters.

\smallskip\noindent\textbf{Duration Predictor}. The duration predictor consists of two convolutional blocks. Each block consists of a 1D time-channel separable convolution, a 1×1 step-wise convolution, layer normalization, ReLU, and dropout. A linear layer along with a softplus layer projects the sequence of hidden representations to a sequence of scalars, which are the predicted phoneme durations. 
 
\smallskip\noindent\textbf{Mel-spectrogram Generator}. To have a lightweight architecture, we use the same variational autoencoder (VAE) based synthesizer as proposed in PortaSpeech~\cite{ren2021portaspeech}.  The synthesizer is comprised of an encoder, a volume-preserving (VP) flow-based prior model and a decoder. The encoder is composed of a 1D-convolution with stride 4 followed by ReLU activation, layer normalization, and a non-causal WaveNet~\cite{oord2016wavenet}. The prior model is a volume-preserving normalizing flow, which is composed of a residual coupling layer and a channel-wise flip operation. The decoder consists of a non-causal WaveNet~\cite{oord2016wavenet} and a 1D transposed convolution with stride 4, also followed by ReLU and layer normalization. 

\section{Training and Evaluation}
\label{sec:experiments}

The system is trained in a multi-step fashion as follows.  To train the image encoder, we employ CE loss to ensure that the encoder transforms the image into hidden representations of phonemes and $\epsilon$, and we adopt the same training configuration as Cai et al.~\cite{cai2021revisiting}.  Next, we freeze the pretrained image encoder and train the expansion module and Mel-spectrogram generator with same configuration as PortaSpeech~\cite{ren2021portaspeech}.  Finally, we train the vocoder with the raw waveform and the groundtruth Mel-spectrogram. Thus, the vocoder's training is independent. We follow HiFiGAN~\cite{kong2020hifi}'s configuration for HiFiGANv2 whose model only contains 0.9M parameters.

\subsection{Training Dataset and Preprocessing}
\label{ssec:training_dataset}
The size of input image is set to 64x224.
Since we use data sets that are designed for ITT tasks and only offer image-word pairs, we utilize a grapheme-to-phoneme (G2P) tool~\cite{g2pE2019} to translate words into phoneme sequences. 
To ensure that our results are reproducible, we use Microsoft\textregistered Azure speech service to synthesize the ground truth audio.
Since we adopt a multi-step training process, we introduce the training data for each step.

\smallskip\noindent\textbf{Image Encoder}. We combine two datasets together: MJSynth (MJ)~\cite{Jaderberg14c}, which contains 9 million word box images generated from a lexicon of 90K English words and Synthtext\cite{Gupta16}. We exclude images that contain non-alphabetic characters. 

\smallskip\noindent\textbf{Sequence Expansion Module and Mel-spectrogram Generator}. We generate the ground truth audio for each word in MJ. We sample 20\% of the images of each word from the cleaned MJ~\cite{Jaderberg14c}. Moreover, to enlarge the sample size of short words, we sample 100 images from Synthtext~\cite{Gupta16} for every word that has less than 6 characters. To increase the speed variation of each word, we randomly apply speed perturbation to each image-audio pair. Empirically, this operation reduces the impact of misalignment caused by the external aligner. We transform the raw waveform with a sampling rate of 22050 Hz into Mel-spectrograms with  frame size 1024 and hop size 256. The dataset contains both image and word information so that we can train ITS and TTS models.

\smallskip\noindent\textbf{Vocoder}. We generate ground truth audio for each utterance in LJSpeech~\cite{ljspeech17} and use that synthesized audio as training data.

\subsection{Testing Dataset}
\label{ssec:testing_dataset}
To tailor the end-to-end ITS system for likely displayed content on personal computing devices we synthesize the top-3000-frequent words based on the word frequency of Wikipedia though an open-source toolkit~\footnote{github.com/Belval/TextRecognitionDataGenerator} with random fonts and color combinations. 

\subsection{Evaluation Metrics}
\label{ssec:Evaluation_Metrics}
Empirically, we found that the Microsoft\textregistered Azure automatic speech recognition (ASR) service performed well on the ground truth audio recordings. Thus, we used the service to transcribe our synthesized output audio. However, since each synthesized output only contains one word and the ASR cannot distinguish among homonyms without context, our evaluations are based on phoneme sequences instead of character sequences. We adopt two evaluation metrics: phone error rate (PER) and word accuracy. 

\section{Experiments}
\label{sec:results}
\subsection{Quality of Audio Synthesis}
\label{ssec:quality_eval}
We compare the performance between our E2E ITS model and a non-E2E ITS pipeline.
A potential advantage of an E2E architecture is reducing the number of parameters. In the standard non-E2E ITS pipeline, there are two encoders, an image encoder for text recognition and a linguistic encoder for Mel-spectrogram synthesis, and a g2p translator.
Our E2E model encodes an image for Mel-spectrogram synthesis with a single image encoder.

We build a non-E2E ITS pipeline as the baseline. We deploy the same architecture as the image encoder in ITS to train an ITT model with MJSynth~\cite{Jaderberg14c} and SynthText~\cite{Gupta16}. The VAE-based TTS~\cite{natspeech} is trained on the same dataset as the E2E ITS model. Table~\ref{tab:exp} presents the PER, accuracy and the size of parameters. 
The TTS system adopts an embedding layer at the bottom to convert input phoneme indices into phoneme embeddings. A phoneme's embedding is universal. On the other hand, our ITS system takes the image encoder's hidden representation to represent phonemes and this representation is affected by variations related to the input image. Based on this difference, we assume the phoneme representation is not as robust as the embeddings. One of our future works will evaluate how the robustness of representation impacts accuracy.

\subsection{Inference Speed}
Since we deploy non-autoregressive models for fast inference speed, we compare the speed of our model with the non-E2E pipeline used in Section~\ref{ssec:quality_eval}. 
We evaluated the inference speed with both recognition speed (images per second) and real-time factor (RTF). The former is used in the ITT field and the latter is used in the TTS field. Since the ITS pipeline has two DNNs, we consider the summation of both DNNs' inference time of a given input as the pipeline's inference time. 
We utilize an Nvidia 2080Ti and synthesize Mel-spectrograms with batch size 1. 
Table~\ref{tab:my_label} shows that the E2E ITS is faster than the ITS pipeline.

\subsection{Impact of Data Distribution}
Today's TTS research focuses on evaluating sentence-level performance~\cite{ren2019fastspeech,ren2020fastspeech,ren2021portaspeech,donahue2020end,elias2021parallel2} as well as the robustness on long utterances~\cite{badlani2022one}.
Clarity of isolated words is rarely studied. 
However, since we are interested in deploying the ITS system for reading isolated words, clear pronunciation is important.
We noticed that both ITS and the VAE-based TTS models perform better when synthesizing words with more phonemes.  
When we sample similar amounts of images for every word from MJSynth~\cite{Jaderberg14c}, the total sample size of 6+ phoneme words is higher. We think the sample distribution over phoneme lengths is an essential factor to the model performance.

\begin{figure}[t!]
    \centering
    \includegraphics[width=\linewidth]{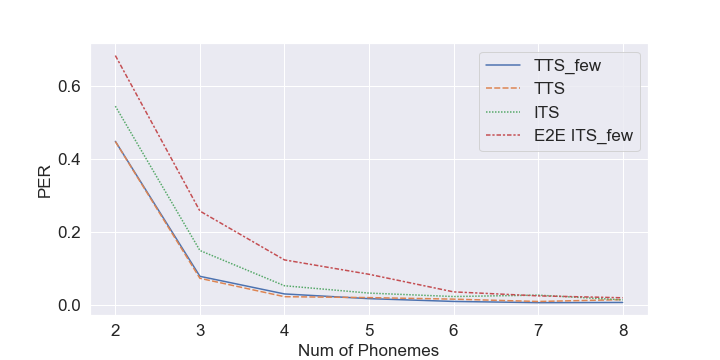}
    \caption{We categorize all testing samples based on their phoneme length and present the model performance on phoneme lengths between 2 and 8.}
    \label{fig:data_size_impact}
\end{figure}

To quantitatively evaluate this impact, we trained new ITS and TTS models utilizing a training set with fewer short-word samples and called these sets \textit{E2E ITS\_few} and \textit{TTS\_few} respectively. We removed all samples that were drawn from Synthtext~\cite{Gupta16} which are short words. 
Figure~\ref{fig:data_size_impact} shows that E2E ITS trained with more short word samples performs noticeably better than its counterpart while the performance on longer words are slightly impacted. 
Moreover, ITS gains more benefit from additional short training samples than TTS.

\begin{table}[]
    \centering
    \begin{tabular}{|c|c|c|c|}
    \hline
    Model Name & PER (\%) & Acc (\%) & Param\\
    \hline
    \hline
    E2E ITS   & 4.7 & 87.8 & 6.1M \\
    \hline
    Non-E2E ITS pipeline & 2.7 & 92.3 & 7.5M \\
    \hline
    \end{tabular}
    \caption{Phone error rate and word accuracy for the image-to-speech systems. Vocoder is not included.}
    \label{tab:exp}
\end{table}

\begin{table}[t!]
    \centering
    \begin{tabular}{|c|c|c|}
    \hline
    Model Name & Speed(image/sec) & RTF \\
    \hline
    \hline
    E2E ITS & 78.0 & 0.0167 \\
    \hline
    Non-E2E ITS pipeline & 59.9 & 0.0236 \\
    \hline
    \end{tabular}
    \caption{Inference speed comparison.}
    \label{tab:my_label}
\end{table}

\section{Conclusion}
\label{sec:conclusion}
We propose the first non-autoregressive E2E ITS system by introducing an intuitive sequence expansion module. This module is flexible enough to accommodate a wide variety of image encoders and Mel-spectrogram generators. Our ITS model is small enough to deploy on mobile devices. Moreover, we demonstrate the importance of data distribution to training ITS models. 
We will enhance the generality of the image-to-phoneme representations in the future.

\bibliographystyle{IEEEbib}
\bibliography{refs}

\end{document}